\documentclass[amsmath,amssymb,
aps, prx,
reprint,
citeautoscript, superscriptaddress]{revtex4-2} 

\usepackage{bbm}
\usepackage{mathrsfs}
\usepackage{bm}
\usepackage{gensymb}
\usepackage{amssymb}
\usepackage{amsfonts}
\usepackage{cancel}
\usepackage{mathtools}
\usepackage{tabularx}

\usepackage{caption}
\usepackage[singlelinecheck=false,justification=raggedright]{caption}
\usepackage{float}
\usepackage{subcaption}
\usepackage[singlelinecheck=false,justification=raggedright]{subcaption}
\usepackage{graphicx}
\usepackage{dcolumn}
\usepackage{placeins}
\usepackage[colorlinks,linkcolor=red,anchorcolor=green,
citecolor=blue,breaklinks]{hyperref}

\usepackage{multirow}
\usepackage{array}

\usepackage{mathrsfs}
\usepackage{color}
\usepackage{braket}
\usepackage[normalem]{ulem}

\graphicspath{{figures/}}
\newcolumntype{C}[1]{>{\centering\arraybackslash}p{#1}}

\pdfoutput=1

\begin{document}

\title{Quantum Spectroscopy with Biphotons: Lyapunov-Based Input–Output Dynamics}

\author{Sameer Dambal}
\affiliation{Department of Physics, 
University of Houston, 
Houston, Texas 77204, United~States}
\email{sadambal@central.uh.edu}

\author{Ajay Ram Srimath Kandada}
\affiliation{Department of Physics and Center for Functional Materials, 
Wake Forest University, 1834 Wake Forest Road, Winston-Salem,
NC 27109, United States}

\author{Eric~R.~Bittner}
\email{ebittner@central.uh.edu}
\affiliation{Department of Physics, University of Houston, Houston, Texas 77204, United~States}

\begin{abstract}
We develop a Lyapunov-based framework to model the evolution of entangled biphotons interacting with cavity and material modes. Using Gaussian-preserving dynamics and M{\o}ller operators, we map input joint spectral amplitudes to experimentally measurable joint spectral intensities. Our model reproduces key features of observed spectra and reveals off-diagonal correlations arising from cavity decay, providing a scalable and tractable tool for quantum spectroscopic analysis.
%
\end{abstract}

\date{\today}
 
\maketitle

\section{Introduction}

Recent advances in quantum light spectroscopy have revealed that quantum correlations in light can serve as powerful probes of electronic and vibrational dynamics in condensed-phase systems~\cite{mukamel2020roadmap, asban2021distinguishability, schlawin2013two, dorfman2016nonlinear, gu2020manipulating, roslyak2009nonlinear, schlawin2012manipulation, schlawin2013suppression, upton2013optically}. By operating in the few-photon regime, these techniques minimize sample perturbation and access information encoded in many-body correlations—features that are often masked in classical measurements. This sensitivity becomes essential when probing ultrafast energy conversion processes in photovoltaic materials or fragile quantum coherence in biological and molecular systems.

A central thrust in this field now focuses on how entangled photons evolve upon interacting with complex matter. This interaction imprints signatures of the material's internal correlations onto the joint properties of the light field~\cite{bittner2020probing, li2017probing, li2019photon, debnath2022entangled, cuevas2018first}. Specifically, spectroscopic protocols that employ time-frequency entangled photon pairs generated via Spontaneous Parametric Down Conversion (SPDC) offer new pathways to interrogate material dynamics with high spectral and temporal resolution~\cite{malatesta2023optical, moretti2023measurement}. In SPDC, a single pump photon generates a correlated pair—signal and idler—distributed over a continuum of frequencies while conserving energy and momentum. The two-photon state is characterized by the joint spectral amplitude (JSA), which encodes both amplitude and phase correlations between the pair.

In practice, time-resolved coincidence counting introduces integration over the detection delay, effectively collapsing the measurement to the joint spectral intensity (JSI)—the squared modulus of the JSA. This object, accessible in experiment, captures the frequency-domain correlations shaped by both the quantum state of the probe and its interaction with the material system. As we demonstrate below, the structure of the JSI encodes key features of the underlying light-matter dynamics and provides a versatile platform for extracting coherence and correlation information beyond the reach of classical spectroscopy.

To overcome the intrinsically low probability of light-matter interactions at the single-photon level, quantum spectroscopic protocols often require operation in the strong coupling regime. This regime is realized by embedding materials within optical microcavities, which enhance the interaction strength by both amplifying the local electric fields and quantizing the electromagnetic (EM) modes that couple to material excitations. These discrete EM modes not only mediate efficient coupling between light and matter, but also enable access to dynamics involving multiple excitons, thereby revealing the structure of many-body correlations that govern material response.

When the coupling strength exceeds the dissipative losses, light and matter hybridize to form polaritons—quasiparticles that inherit characteristics of both photons and material excitations. Polaritons serve as a robust platform for exploring collective quantum phenomena, including Bose-Einstein condensation~\cite{deng2010exciton, kasprzak2006bose, keeling2020bose}, superfluid transport and vortex formation~\cite{amo2009superfluidity, lerario2017room}, and nonlinear optical behavior such as bistability~\cite{amthor2015optical, pickup2018optical}, four-wave mixing~\cite{dorfman2021multidimensional}, and soliton propagation~\cite{sich2016soliton, grosso2011soliton}. In suitably engineered photonic lattices, polaritons can acquire topological properties, exhibiting chiral edge modes and disorder-immune transport analogous to the quantum Hall effect and topological insulators~\cite{klembt2018exciton, karzig2015topological}. Furthermore, in two-dimensional van der Waals materials such as $\mathrm{MoO_3}$ and graphite, phonon-polaritons exhibit hyperbolic dispersion~\cite{wang2024planar, fang2020directional, nemilentsau2016anisotropic}, enabling deep subwavelength confinement and directional energy flow.
These strongly coupled light-matter systems present an ideal testbed for quantum spectroscopy. By probing their response with single- and biphoton sources, one can extract signatures of many-body interactions, quantum correlations, and nonclassical dynamics that remain hidden in conventional optical measurements.

Interpreting experimental results in quantum spectroscopy demands a rigorous theoretical framework, and several such approaches have recently emerged. Bittner~\cite{bittner2020probing} demonstrated that biphoton scattering and radiative cascade processes in coupled molecular dimers can generate entanglement entropy in the outgoing biphoton state. Modeling the dimer as a pair of anharmonically interacting excitons embedded in a photonic continuum, they employed diagrammatic techniques to show that the output entropy correlates with both the interaction strength and the repulsive coupling. Complementary work by Li and coworkers~\cite{li2017probing, li2019photon} developed a related framework using the Dicke model and input-output theory~\cite{gardiner1985input, gardiner2004quantum} to analyze photon-photon correlations in a Hong-Ou-Mandel (HOM) configuration~\cite{bouchard2020two}. In this scheme, only one photon from a time-frequency entangled pair interacts with a resonant medium, and its output state is shaped by a transmission function that encodes the system's spectral response. The poles of this function define the underlying polariton branches and reveal the many-body dynamics of the medium.

While these models successfully capture nonlinear photophysics and many-body correlations, they often involve Hilbert spaces that scale exponentially with photon number and mode discretization. For instance, modeling a biphoton interacting with a cavity polariton via the Jaynes-Cummings Hamiltonian yields a Hilbert space dimension of order $\mathcal{O}(2^{2N+2})$, where $N$ is the number of signal and idler photon modes. This exponential growth poses significant challenges for numerical simulation and data interpretation. Moreover, directly mapping the input JSA to the output JSI—the experimentally accessible observable—remains a nontrivial task in many of these models. Developing a reduced, yet physically accurate, framework that captures this mapping while remaining computationally tractable would provide a robust foundation for extending quantum spectroscopy to more complex and strongly correlated systems.

We employ M{\o}ller operators to connect the input, interacting, and output states of the system, thereby eliminating the need for discrete-time gating and establishing a framework directly aligned with experimental practice. These operators provide a natural mechanism for linking the input and output covariance matrices, with the transformation governed by Lyapunov equations~\cite{kinsner2006characterizing}. This formalism enables the direct extraction of key observables, such as the purity of the output biphoton state, from the Wigner function expressed in terms of the covariance matrix. Importantly, the same framework generalizes naturally to the Tavis-Cummings limit, allowing for systematic exploration of collective light-matter interactions in many-body systems. Within this approach, we show that the predicted output JSI  reproduces key features observed in experiments involving biphoton interactions with an empty microcavity, underscoring the consistency and applicability of our model.

\section{Theoretical Framework}

\subsection{Hamiltonian}

Consider a coupled-oscillator model of frequency-entangled bath photons interacting with cavity modes and material excitations according to the following Hamiltonian,
\begin{widetext}
\begin{eqnarray}
    \label{Sys_Hamiltonian}
    H &=& \underbrace{\hbar \omega_c \hat{a}^\dagger\hat{a}}_{\hat{H}_0^{c}} +
    \overbrace{\hbar\int d\omega \omega\hat{b}_{s/i}^\dagger(\omega)\hat{b}_{s/i}(\omega)}^{\hat{H}_0^{s/i}} + \underbrace{\hbar\sum_j\Omega_j\hat{S}_j^\dagger\hat{S}_j^-}_{\hat{H}^m_0} + \overbrace{\underbrace{\hbar g\int d\omega\left(\hat{a}^\dagger\hat{b}_{s/i}(\omega)+\hat{b}_{s/i}^\dagger(\omega)\hat{a}\right)}_{\hat{H}^{s/i}_{int}} - \underbrace{i\hbar\sqrt{\kappa}\sum_j(\hat{a}\hat{S}_j^+ - \hat{a}^\dagger\hat{S}_j^-)}_{\hat{H}^m_{int}}}^{\hat{H}^c_{int}} \nonumber \\ 
\end{eqnarray}
\end{widetext}
where $\hat{a}, \hat{a}^\dagger$ are the cavity modes, $\hat{b}_{s/i}(\omega), \hat{b}_{s/i}^\dagger(\omega)$ are the biphoton signal/idler operators, and the $\hat{S}^\dagger_i, \hat{S}_i$ are operators of the material modes that are assumed to be bosonic. The first 3 terms of Eq. \eqref{Sys_Hamiltonian} are the free Hamiltonians of the cavity, signal/idler photons, and the material, respectively. The fourth term describes the bilinear coupling between the signal/idler modes of the photons, and the last term describes the coupling between the cavity and material modes. The coupling terms obey the first Markov and rotating wave approximations. The former assumes a uniform coupling strength, and the latter removes rapid counter-rotating terms.  

We assume that the signal and idler photons begin in an entangled state given by, 
\begin{eqnarray}
    \label{input_state}
    |\psi\rangle_{in} = \iint d\omega_s d\omega_i {\cal F}_{in}(\omega_s, \omega_i)\hat{b}^\dagger_{in}(\omega_s)\hat{b}^\dagger_{in}(\omega_i)|0\rangle_{in}
\end{eqnarray}
where ${\cal F}_{in}(\omega_s, \omega_i) = _{in}\langle 0|\hat{b}_{in}^\dagger(\omega_i)\hat{b}^\dagger_{in}(\omega_s)|0\rangle_{in}$ is the joint spectral amplitude (JSA) of the input photons. This two-dimensional correlation function quantifies the degree of entanglement between the signal and idler photon states. 

\begin{figure}
    \centering
    \includegraphics[width=1\linewidth]{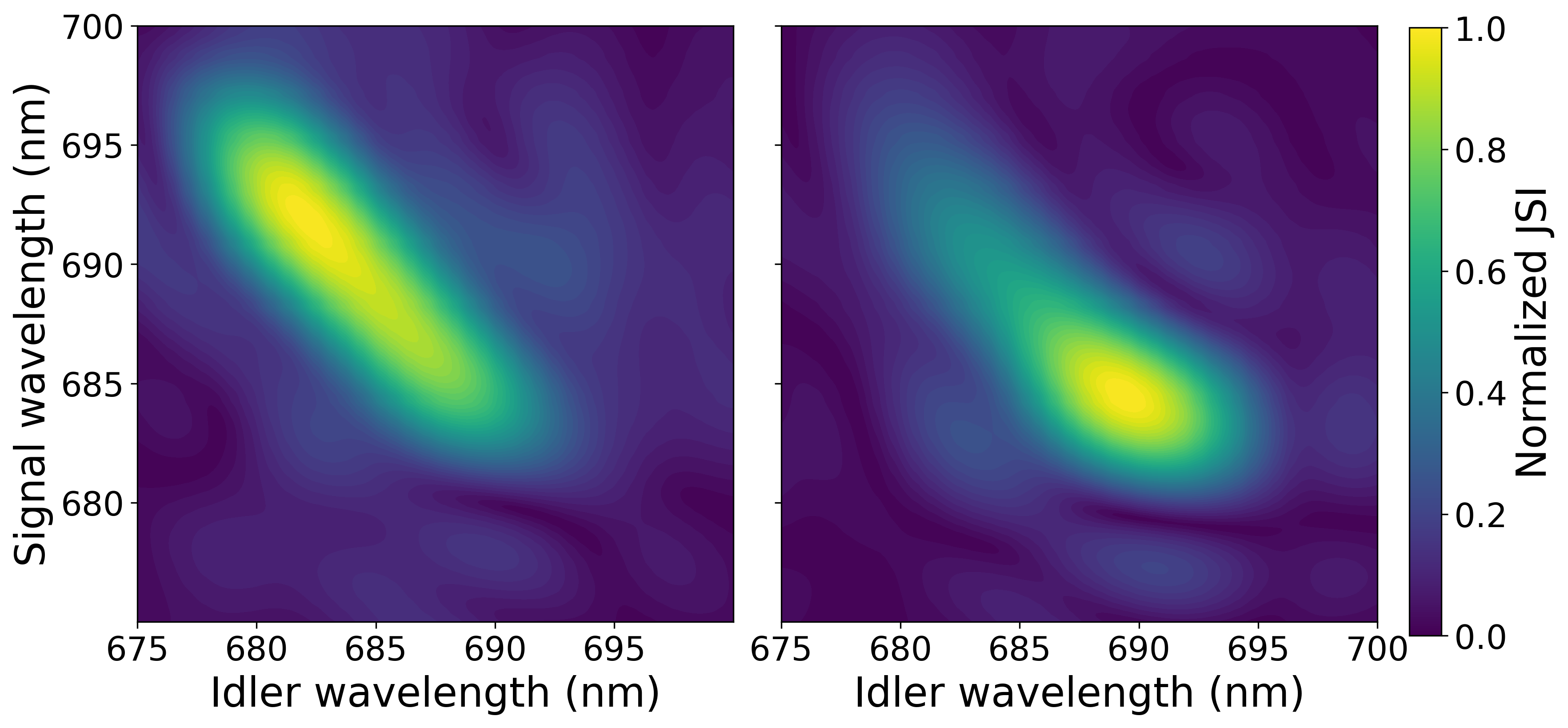}
    \caption{Experimental input and output JSI through an empty microcavity. The input is a normalized state generated using SPDC as described in \citet{malatesta2023optical}. Data reproduced from  \citet{malatesta2023optical}}. 
    \label{expt_JSI_results}
\end{figure}

Fig. \ref{expt_JSI_results} shows an example of an experimental input and output biphoton JSI for empty microcavities. The input JSA represents a spectrally entangled state that is generated in a Type-I $\beta$-Barium Borate (BBO) crystal phase-matched for SPDC with the pump wavelength of $343$ nm \cite{malatesta2023optical}. This dictates the signal/idler dispersion in Eq. \eqref{Sys_Hamiltonian}. Several different input states have also been experimentally produced as demonstrated in \citet{moretti2023measurement}. Similarly, the output state can be defined as,
\begin{align}
    \label{output_state}
    |\psi\rangle_{out} = \iint d\omega_s' d\omega_i' {\cal F}_{out}(\omega_s', \omega_i')\hat{b}^\dagger_{out}(\omega_s')\hat{b}^\dagger_{out}(\omega_i')|0\rangle_{out}
\end{align}
The subscripts ``in'' and ``out'' in the operators $\hat{b}_{in/out}, \hat{b}^\dagger_{in/out}$ act on the input and output vacuum $|0\rangle_{in}, |0\rangle_{out}$ respectively. According to the input-output formalism \cite{gardiner1985input, gardiner2004quantum}, the output and the input operators are related to the coupled cavity mode as,
\begin{eqnarray}
    \label{input_output}
    \hat{b}_{out}(t) &=& \hat{b}_{in}(t) + \sqrt{\kappa}\hat{a}(t)
\end{eqnarray}
This form of the input-output relation assumes the transmission of the input photons completely onto the output without reflections back to the input vacuum.
Since these modes admit asymptotic solutions in time as the initial and final states, we can relate the freely evolving input and output modes from Eqs. \eqref{input_state}, \eqref{output_state} with the interacting modes in Eq. \eqref{Sys_Hamiltonian} using M{\o}ller operators \cite{moller1946new}.
\begin{eqnarray}
    \label{Moller_op}
    \hat{\Omega}_\pm = \lim_{t\rightarrow\mp\infty} e^{iHt}e^{-iH_0t} \\
    \label{bath_Moller}
    \hat{\Omega}_\pm \hat{b}_{in/out}(t) = \hat{b}(t)
\end{eqnarray}
where $H_0$ and $H$ are the free and interacting Hamiltonians, and $\hat{b}_{in/out}(t), \hat{b}(t)$ are the free and interacting modes respectively.  More generally, we can define a vector of modes as,
\begin{eqnarray}
     \vec{x}(t) = (\hat{b}_s(t), \hat{b}_i(t), \hat{a}(t),
\hat{S}(t))^T
\end{eqnarray}
and define the M{\o}ller operators for these modes as
\begin{eqnarray}
    \label{Moller_signal}
    \hat{\Omega}^{s/i}_\pm &=& \lim_{t\rightarrow\mp\infty} e^{i[\hat{H}^{s/i}_0 + \hat{H}^{s/i}_{int}]t} \cdot e^{-i\hat{H}_0^{s/i}t} \\
    \label{Moller_cavity}
    \hat{\Omega}^c_\pm &=& \lim_{t\rightarrow\mp\infty} e^{i[\hat{H}_0^{c} + \hat{H}^{c}_{int}]t}\cdot e^{-i\hat{H}_0^{c}t} \\
    \label{Moller_material}
    \hat{\Omega}^m_\pm &=& \lim_{t\rightarrow\mp\infty} e^{i[\hat{H}_0^m + \hat{H}^m_{int}]t} \cdot e^{-i\hat{H}_0^m t} \\
    \vec{x}(t) &=& \operatorname{diag}(\hat{\Omega}^s_\pm, \hat{\Omega}^i_\pm, \hat{\Omega}^c_\pm, \hat{\Omega}^m_\pm) \vec{x}_{in/out}
\end{eqnarray}
where the specified Hamiltonians are marked in Eq. \eqref{Sys_Hamiltonian}. 

\subsection{Governing Equation of the Model}
The Hamiltonian in Eq. \eqref{Sys_Hamiltonian} can be numerically intractable with current computational capabilities. To make this problem solvable, we note that all the operators in the Hamiltonian are bilinear. As a result, the unitary operator $e^{-iHt}$ is a Gaussian-preserving map, allowing us to reflect the dynamics of our system in the first moments and covariances \cite{ferraro2005gaussian}. We can also use Wick's theorem to contract higher-order moments into these first- and second-order moments. This is useful since we can perform our dynamics with finite-matrix theory without dealing with an exponentially growing Hilbert space. The equation of motion for the operators can then be written in a linearized form as,
\begin{eqnarray}
    \label{x_eom_0}
    \frac{d\mathbf{x}}{dt} &=& -i[H,\mathbf{x}] + {\cal L}(\mathbf{x}) \\
    \label{x_eom}
    &=&  W\cdot\mathbf{x}
\end{eqnarray}
where ${\cal L}$ is the Lindbladian and $W$ is the dynamical matrix derived from the Heisenberg equations of motion. Assuming a high Q-factor of the cavity, we shall ignore the dissipative dynamics, and consequently the Lindbladian from Eq. \eqref{x_eom_0} in this study. For one pair of signal and idler photons coupled to the cavity and material excitation, the dynamical matrix, $W$, takes the form,
\begin{eqnarray}
    \label{dynamical_matrix}
    W &=& \begin{pmatrix}
        -i\omega_s & 0 & -ig & 0 \\
        0 & -i\omega_i & -ig & 0 \\
        -ig & -ig & -i\omega_c & -\sqrt{\kappa} \\
        0 & 0 & -\sqrt{\kappa} & -i\Omega
    \end{pmatrix}
\end{eqnarray}

For the second moments, we can define the covariance matrix as,
\begin{eqnarray}
    \Theta = \langle x \cdot x^\dagger\rangle - \langle x \rangle \cdot \langle x^\dagger \rangle
\end{eqnarray}
and the equation of motion for the covariances can be derived to be,
\begin{eqnarray}
    \frac{d\Theta(t)}{dt} &=& \left\langle \frac{dx}{dt} \cdot x^\dagger + x\cdot \frac{dx^\dagger}{dt} \right\rangle \nonumber \\
    &-& \left\langle \frac{dx}{dt} \right\rangle \langle x^\dagger \rangle -\langle x\rangle \left\langle \frac{dx^\dagger}{dt} \right\rangle\\
    \label{Sylvester_diff_eqn}
    \frac{d\Theta(t)}{dt} &=& W \cdot \Theta(t) + \Theta(t)\cdot W^\dagger 
\end{eqnarray} 
where Eq. \eqref{x_eom} has been used. Eq. \eqref{Sylvester_diff_eqn} is known as the Sylvester differential equation, and its stationary form is the well-known Lyapunov equation used in the study of chaos \cite{kinsner2006characterizing}. Solving the Sylvester equation allows us to understand the evolution of signal-idler correlations as they interact with a microcavity. In solving this, we must ensure that it is consistent with the boundary conditions of the correlation matrix. To establish this, we convert it into its respective input and output forms as follows:
\begin{eqnarray}
    \label{input_sylvester}
    \frac{d\Theta_{in}(t)}{dt} &=& \Theta_{in}(t)\cdot W^\dagger + W \cdot \Theta_{in}(t) \\
    \label{output_sylvester}
    \frac{d\Theta_{out}(t)}{dt} &=& \Theta_{out}(t)\cdot W + W^\dagger \cdot \Theta_{out}(t)
\end{eqnarray}
Since the output correlation matrix evolves back in time, we reflect that by switching the Hermitian conjugated dynamical matrices in the two terms. Taking the Laplace transform of Eqs. \eqref{input_sylvester} and \eqref{output_sylvester}, we get,
\begin{eqnarray}
    z\tilde{\Theta}_{in}(z) - \Theta_{in}(t\rightarrow-\infty) &=& \tilde{\Theta}_{in}(z)\cdot W^\dagger + W \cdot \tilde{\Theta}_{in}(z) \nonumber \\ \\
    z\tilde{\Theta}_{out}(z) - \Theta_{out}(t\rightarrow+\infty) &=& \tilde{\Theta}_{out}(z)\cdot W + W^\dagger \cdot \tilde{\Theta}_{out}(z) \nonumber \\
\end{eqnarray}
where $\Theta_{out}(t\rightarrow+\infty)$ is the correlation matrix of the output. We 
take $z=0$ so that $\tilde{\Theta}(z=0) = \int \Theta(t)dt$ is the integration of correlations over all time. Since the biphotons are taken to be non-interacting at asymptotic times and interacting at finite times, this time-integrated quantity effectively captures the experiment's dynamics. It evolves under the full Hamiltonian and is useful in setting up our boundary conditions. Fixing $z=0$, we obtain, 
\begin{eqnarray}
     \tilde{\Theta}_{in}(0)\cdot W^\dagger + W \cdot \tilde{\Theta}_{in}(0) + \Theta_{in}(t\rightarrow-\infty) = 0 \\
    \tilde{\Theta}_{out}(0)\cdot W + W^\dagger \cdot \tilde{\Theta}_{out}(0) + \Theta_{out}(t\rightarrow+\infty) = 0
\end{eqnarray}

Solving for these equations, we get,
\begin{eqnarray}
    \label{Boundary_condition}
    && \Theta_{out}(t\rightarrow+\infty) = \tilde{\Theta}_{in}(z=0)\cdot W^\dagger + W \cdot \tilde{\Theta}_{in}(z=0) \nonumber \\
    &-& \tilde{\Theta}_{out}(z=0)\cdot W - W^\dagger \cdot \tilde{\Theta}_{out}(z=0) + \Theta_{in}(t\rightarrow-\infty) \nonumber \\
\end{eqnarray}

In order to find the output correlation function after all interactions have died out, we need to express $\tilde{\Theta}_{out}(z=0)$ in terms of $\tilde{\Theta}_{in}(z=0)$. For this, we go back to Eq. \eqref{x_eom} and take its Laplace transform,
\begin{eqnarray}
    \label{eom_in}
    z.\tilde{x}(z) - x_{in}(t \rightarrow -\infty) = W_{in}\tilde{x}(z)
\end{eqnarray}
where $W_{in} = W$ is the dynamical matrix responsible for propagating the vector of moments forward in time from $t \rightarrow -\infty$. To connect this to the moments at $t \rightarrow +\infty$, we define $W_{out} = W^\dagger$ and propagate the moments backward in time using the Laplace transform,
\begin{eqnarray}
    \label{eom_out}
    z.\tilde{x}(z) - x_{out}(t \rightarrow +\infty) = W_{out}\tilde{x}(z)
\end{eqnarray}
where we can eliminate the intermediate vector of operators and obtain,
\begin{eqnarray}
    \label{in_out_connection}
    x_{out}(z) = -(W_{out} - z)(W_{in} - z)^{-1}x_{in}(z)
\end{eqnarray}

Let us assign $(W_{out} - z)(W_{in} - z)^{-1} = S$. This matrix takes the form of M{\o}ller operators as defined in Eqs. \eqref{Moller_signal} - \eqref{Moller_material}. Eq. \eqref{in_out_connection} physically says that the input modes are propagated forward in time and the output modes are propagated backward in time under the interacting Hamiltonian. In doing so, both modes must meet at a common point in time, $t$, to maintain continuity. This forms the essence of the connection between the input and output Heisenberg operators. Since the covariance matrix is defined as,
\begin{eqnarray}
    \tilde{x}(z)_{in/out} \cdot \tilde{x}(z)^\dagger_{in/out} = \tilde{\Theta}_{in/out}(z)
\end{eqnarray}
we can write the transformation between the input and output covariance matrices as,
\begin{eqnarray}
    \label{out_to_in}
    \tilde{\Theta}_{out}(z=0) = -S\tilde{\Theta}_{in}(z=0)S^\dagger
\end{eqnarray}

Now we substitute Eq. \eqref{out_to_in} in \eqref{Boundary_condition} to obtain the connection between the input and output covariance matrices as, 
\begin{eqnarray}
    \label{in_out_corr_connection}
    &&\Theta_{out}(t\rightarrow +\infty) = \tilde{\Theta}_{in}(z=0). W^\dagger + W \cdot \tilde{\Theta}_{in}(z=0) \nonumber \\
    &+& S \tilde{\Theta}_{in}(z=0) S^\dagger \cdot W + W^\dagger \cdot S \tilde{\Theta}_{in}(z=0)S^\dagger \nonumber \\ 
    &+& \Theta_{in}(t\rightarrow -\infty)
\end{eqnarray}
Thus, Eq. \eqref{in_out_corr_connection} establishes the connection between the input correlations at $t \rightarrow -\infty$ with the output correlations at $t \rightarrow +\infty$ and becomes the governing equation of our method.  

In the absence of the signal/idler photon interactions with the cavity, the dynamical matrix represents a free-evolution, $W = W_0$, where the off-diagonal signal/idler-cavity coupling $g=0$. The signal/idler subspace of $W_0$ is then diagonal. Consequently, these photons do not interact with the material degrees of freedom. In this case, the first four terms of Eq. \eqref{in_out_corr_connection} do not contribute to the signal/idler subspace of the output covariance matrix, $\Theta_{out}(t\rightarrow+\infty)$ and thus, the output JSA is identical to the input JSA contained in the initial covariance matrix, $\Theta_{in}(t\rightarrow-\infty)$. When $g\neq0$, the first four terms make a non-diagonal contribution to the subspace of $\Theta_{out}(t\rightarrow +\infty)$, and we obtain a mapping from $\Theta_{in}(t\rightarrow -\infty) \rightarrow \Theta_{out}(t\rightarrow +\infty)$ under interactions.  

\begin{figure*}[!ht]
    \centering
    \begin{subfigure}{0.95\linewidth}
        \centering
        \includegraphics[width=0.75\linewidth]{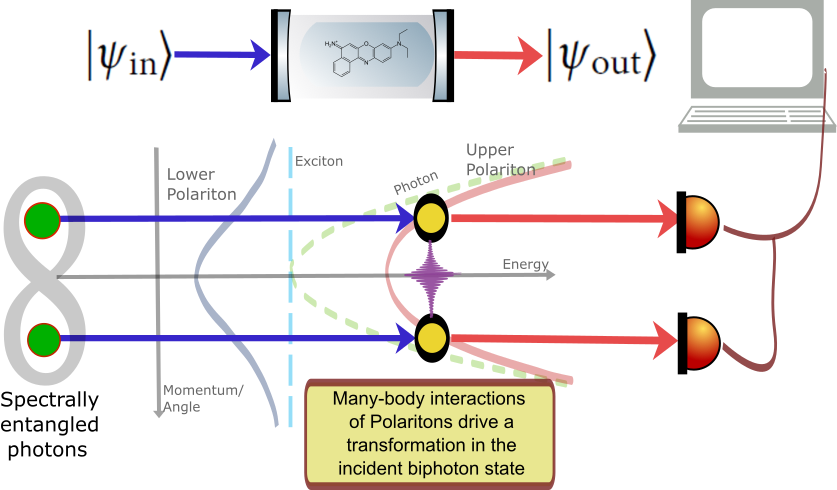}
        \caption{}
        \label{Better_eye_candy}
    \end{subfigure}
    
    \begin{subfigure}{0.95\linewidth}
        \centering
        \includegraphics[width=0.95\linewidth]{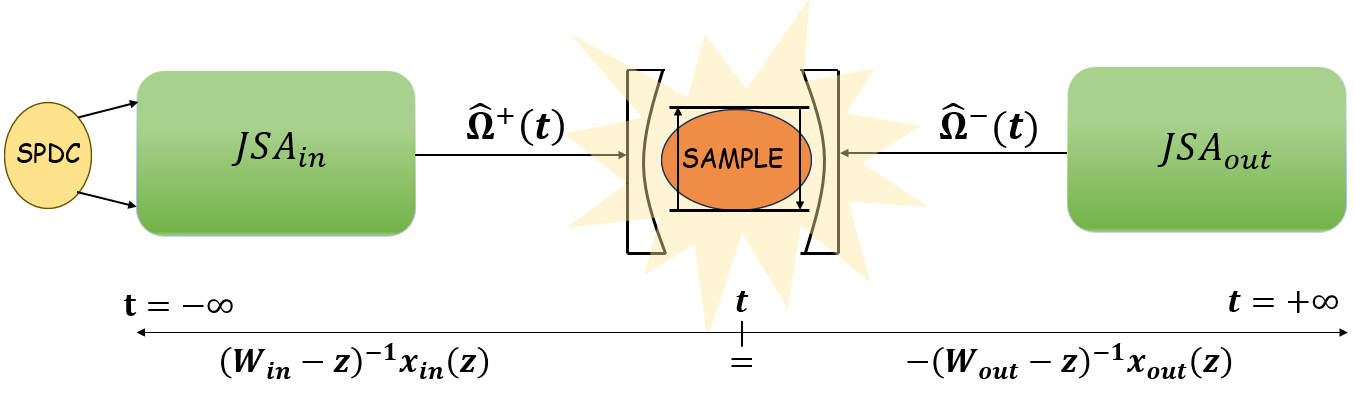}
        \caption{}
        \label{Eye_candy}
    \end{subfigure}
    \caption{Broad overview of the spectroscopic apparatus and the theoretical framework. (a) Schematic of the experimental setup demonstrated by the theoretical framework developed in this work. (b) Schematic of the boundary conditions connecting the input to the output mediated by interactions with a cavity.}
    \label{fig:combined_schematic}
\end{figure*}

This model is also robust to the number of material degrees of freedom present in the cavity as long as the Hamiltonian in Eq. \eqref{Sys_Hamiltonian} generates a Gaussian-preserving map. This means that it can be applied equally well to monomeric, dimeric, and other polymeric systems under the two-level approximation. An illustration of such an experimental setup and the theoretical boundary conditions that we described earlier is shown in Figs. \ref{Better_eye_candy} and \ref{Eye_candy}. 

\subsection{Extension to continuous spectrum}

Having laid the theoretical groundwork to describe the experiment, we can now trivially extend this model and create a continuous spectrum of frequency-entangled biphotons. Recognizing that the JSA is a correlation between the signal and idler frequencies, we enter this into the off-diagonal elements of the signal/idler subspace in the extended covariance matrix. The extended dynamical and covariance matrices now look like,

\begin{widetext}
\begin{eqnarray}
    \label{dynamical_matrix_continuum}
    W &=& \begin{pmatrix}
    - i \omega_{s1} & 0 & \cdots & 0 & 0 & 0 & \cdots & 0 & -ig & 0\\
    0 & - i \omega_{s2} & \cdots & 0 & 0 & 0 & \cdots & 0  & -ig & 0\\
    \vdots & \vdots & \ddots & \vdots & \vdots&\vdots & \ddots & \vdots &\vdots & \vdots \\
    0 & 0 & \cdots & - i \omega_{s n} & 0 & 0 & \cdots & 0 & -ig & 0 \\
    0 & 0 & \cdots & 0 & -i\omega_{i1} & 0 & \cdots & 0 & -ig & 0 \\
    0 & 0 & \cdots & 0 & 0 & - i \omega_{i2} & \cdots & 0 & -ig & 0 \\
    \vdots& \vdots & \ddots & \vdots & \vdots & \vdots & \ddots & \vdots & \vdots & \vdots \\
    0 & 0 & \cdots & 0 & 0 & 0 & \cdots & -i \omega_{in} & -ig & 0\\
    -ig & -ig & \cdots & -ig & -ig & -ig & \cdots & -ig & -i\omega_c & -\sqrt{\kappa} \\
    0 & 0 & \cdots & 0 & 0 & 0 & \cdots & 0 & -\sqrt{\kappa} & -i\Omega
    \end{pmatrix} \\
    \label{cov_matrix_continuum}
    \Theta_{in}(t\rightarrow -\infty) &=& \begin{pmatrix}
        \frac{1}{2} & 0 & \cdots & 0 & F(\omega_{s1},\omega_{i1})& F(\omega_{s1},\omega_{i2}) & \cdots & F(\omega_{s1},\omega_{in}) & 0 & 0 \\
        0 & \frac{1}{2} & \cdots & 0 & F(\omega_{s2},\omega_{i1}) & \vdots & \cdots & F(\omega_{s2},\omega_{in}) & 0 & 0 \\
        \vdots & \vdots & \ddots & 0 & \vdots & \vdots& \ddots & \vdots & \vdots & \vdots\\
        0 & 0 & \cdots & \frac{1}{2} & F(\omega_{sn},\omega_{i1}) &F(\omega_{sn},\omega_{i2}) & \cdots & F(\omega_{sn},\omega_{in}) & 0 & 0 \\
        F(\omega_{s1},\omega_{i1})& F(\omega_{s1},\omega_{i2}) & \cdots & F(\omega_{s1},\omega_{in}) & \frac{1}{2} & 0 & \cdots & 0 & 0 & 0 \\
        F(\omega_{s2},\omega_{i1}) & \vdots & \cdots & F(\omega_{s2},\omega_{in}) & 0 & \frac{1}{2} & \cdots & 0 & 0 \\
        \vdots & \vdots& \ddots & \vdots & \vdots & \vdots & \ddots & \vdots & \vdots & \vdots \\
        F(\omega_{sn},\omega_{i1}) & F(\omega_{sn},\omega_{i2}) & \cdots & F(\omega_{sn},\omega_{in}) & 0 & 0 & \cdots & \frac{1}{2} & 0 & 0 \\
        0 & 0 & \cdots & 0 & 0 & 0 & \cdots & 0 & \frac{1}{2} & 0 \\
        0 & 0 & \cdots & 0 & 0 & 0 & \cdots & 0 & 0 & \frac{1}{2}
    \end{pmatrix}
\end{eqnarray}    
\end{widetext}
These matrices can now be input in Eq. \eqref{in_out_corr_connection} to obtain the output biphoton correlations.

\subsection{Entanglement Entropy}

To find the entanglement of the output biphoton states, we perform a Schmidt decomposition of the output JSA.
\begin{eqnarray}
    \label{svd}
    {\cal F}_{out}(\omega_1, \omega_2) = \sum_n r_n U_n(\omega_1)V_n(\omega_2)
\end{eqnarray}
where $r_n$ are the singular values and $U_n(\omega_1), V_n(\omega_2)$ are the left and right eigenvalues of the JSA. The von Neumann entropy can then be calculated as,
\begin{eqnarray}
    \label{vn-entropy}
    S = -\sum_n r_n^2 ln r_n^2
\end{eqnarray}

To derive the purity of the state, we Weyl-transform the problem and express the corresponding Wigner function of our multi-mode state in terms of the covariance matrix as described in Ref \cite{ferraro2005gaussian}. This takes the form, 
\begin{eqnarray}
    \label{Wigner_function}
    {\cal W}[\rho](\alpha) = \frac{e^{-\frac{1}{2}(\alpha-\bar{\alpha})^T \Theta^{-1} (\alpha-\bar{\alpha})}}{(2\pi)^n \sqrt{|\Theta|}}
\end{eqnarray}

Here, $\alpha \in \mathbb{C}^{2m+2n}$, $m,n$ are the dimensions of the bipartitioned subspaces. We take $\bar{\alpha} = 0$ since local transformations do not change the entanglement structure. Since purity is defined as $\mu = Tr(\rho^2)$, we observe that for any operator $O_k$ that admits a well-defined Wigner function, $W_k(\alpha)$, we can use the overlap property between Wigner functions and write,
\begin{eqnarray}
    \mu(\Theta(t \rightarrow+\infty)) &=& \int d^{2n}X \cdot {\cal W}^2(X) \\
    &=& \frac{1}{\sqrt{|\Theta(t\rightarrow +\infty)|}}
\end{eqnarray}

This gives us the framework to calculate the purity and, subsequently, the mutual information of the output biphoton state. This Wigner function approach aided by the correlation matrix provides a strong theoretical tool to calculate many other observables as described in Ref \cite{ferraro2005gaussian}.

\section{Results}

\subsection{Gaussian Initial JSI}

\begin{figure}
    \centering
    \includegraphics[width=1.0\linewidth]{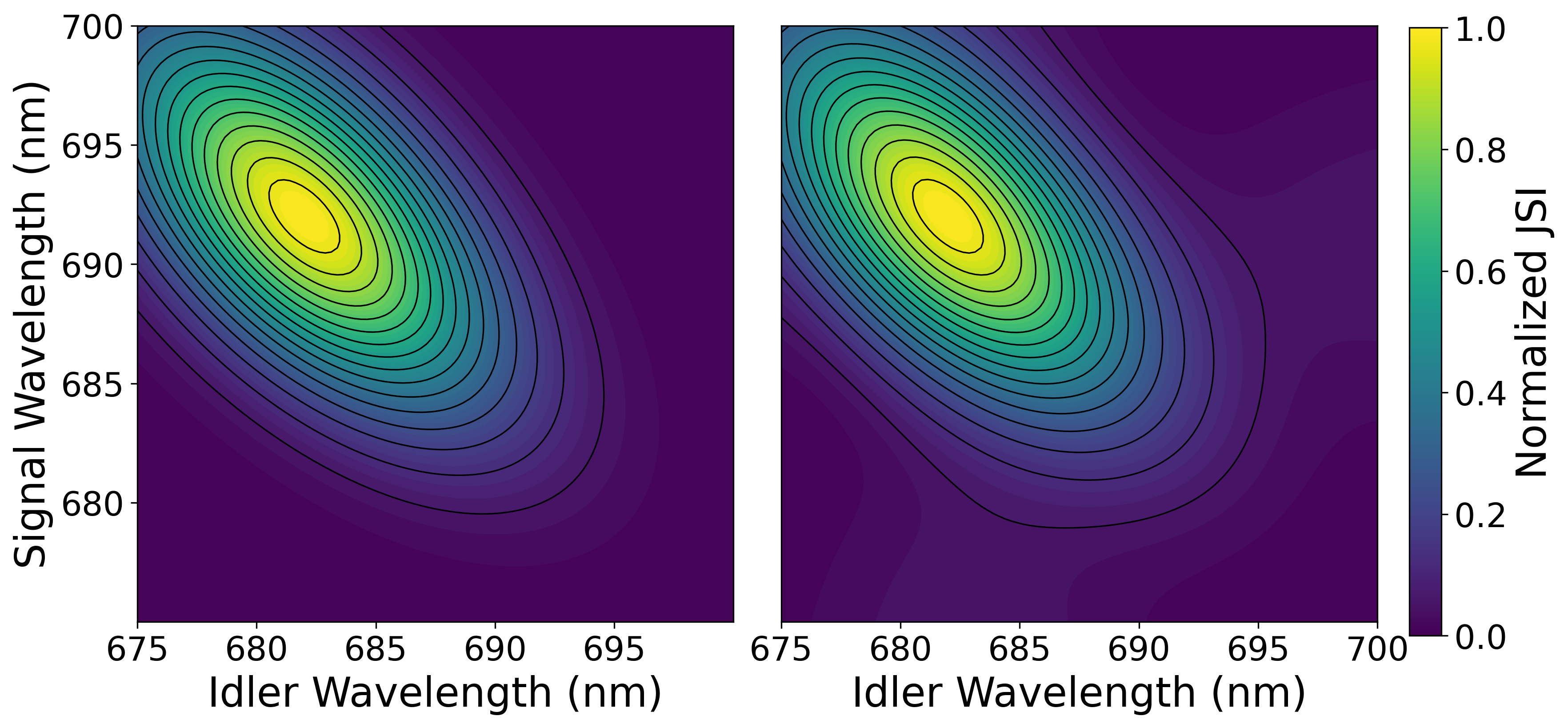}
    \caption{Initial and Final JSI for a theoretical squeezed gaussian. The final JSI is simulated at $\sqrt{\kappa} = 488$ meV ($2540.7$ nm). Other parameters of the model are $\omega_c = \Omega = 1809$ meV ($685.4$ nm), and peak of the squeezed gaussian in the above plot in frequency is at $\omega_i = 1819$ meV ($681.6$ nm), $\omega_s = 1790$ meV ($692.7$ nm).}
    \label{fig:gaussian_JSI}
\end{figure}

To further illustrate our theoretical framework, we begin by considering an input JSI that is a squeezed antisymmetric Gaussian state centered at any arbitrary point along the diagonal. The correlation matrix can be trivially written for this state and the resulting output correlation matrix can be obtained using Eq. \eqref{in_out_corr_connection}. In Fig. \ref{fig:gaussian_JSI}, we observe that this does not cause a significant change in the output JSI. This is because a squeezed Gaussian corresponds to a state of low entropy. As a result, this leaves little room to bring an observable change in the final JSI from any information transfer into the cavity.

\subsection{Experimental Initial JSI}

\begin{figure*}[!htbp]
    \centering
    \begin{subfigure}[b]{0.45\linewidth}
        \centering
        \includegraphics[width=\textwidth]{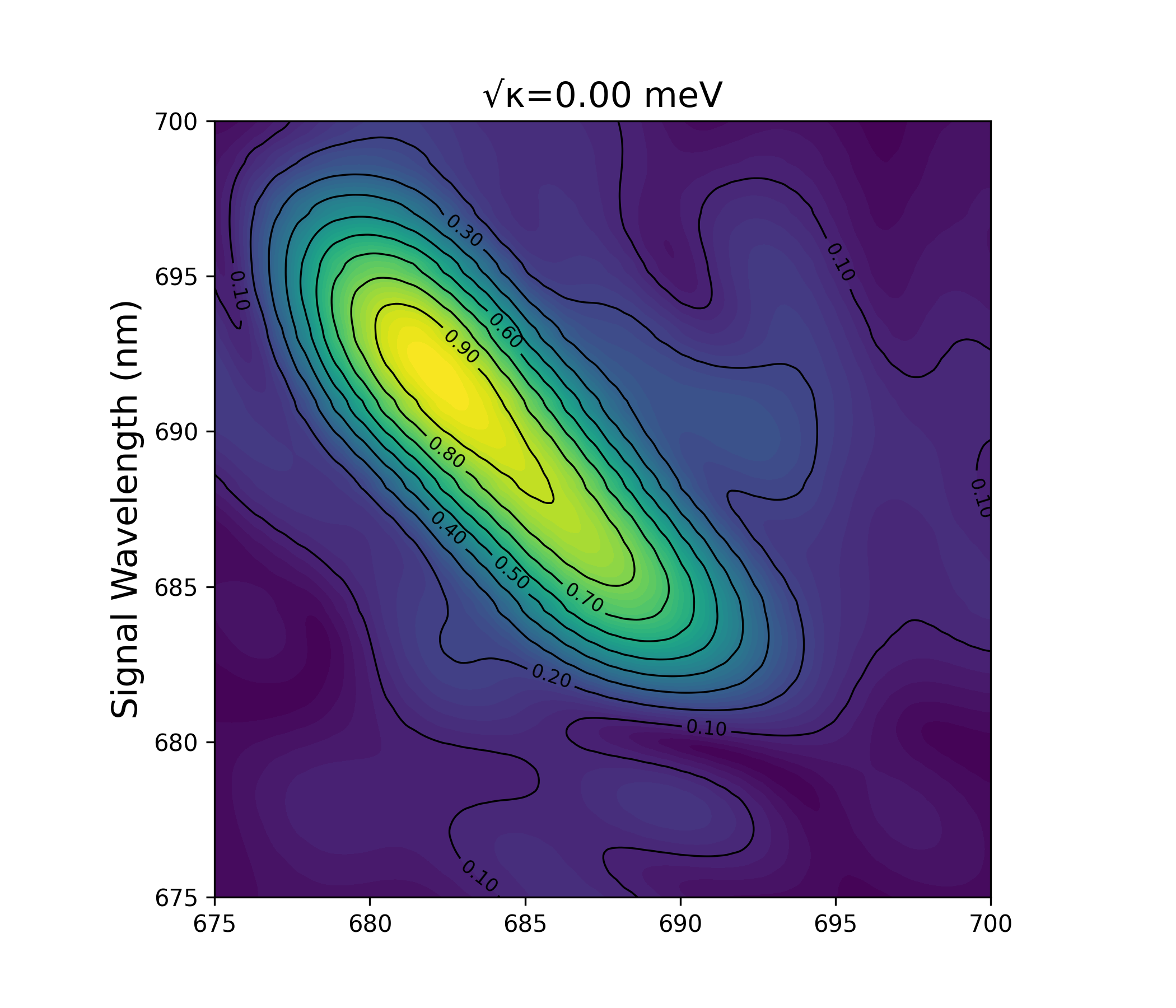}
        \caption{}
        \label{JSI_output_0}
    \end{subfigure}
    \hspace{-0.1\linewidth} 
    \begin{subfigure}[b]{0.45\linewidth}
        \centering
        \includegraphics[width=\textwidth]{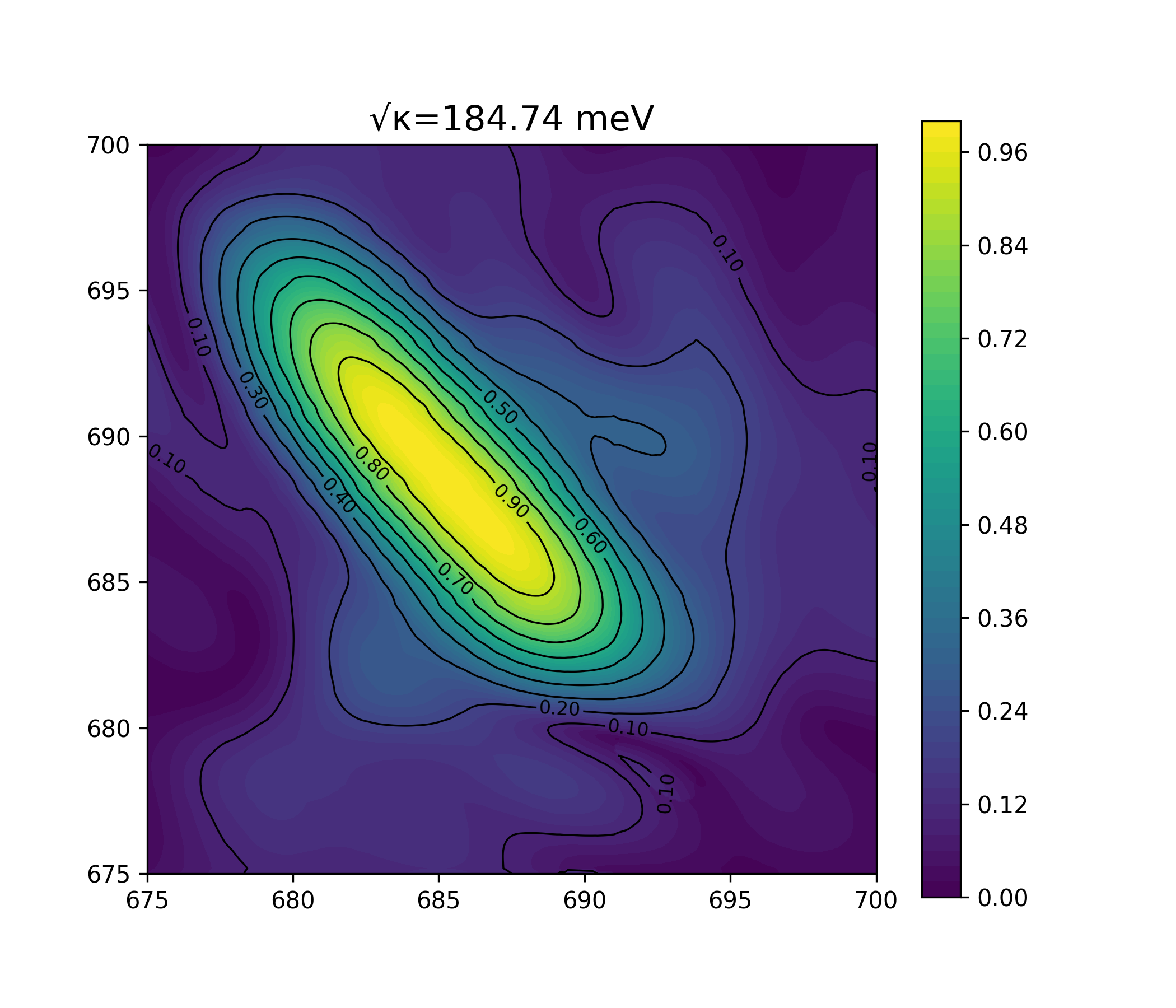}
        \caption{}
        \label{JSI_output_1}
    \end{subfigure}
    
    \begin{subfigure}[b]{0.45\linewidth}
        \centering
        \includegraphics[width=\textwidth]{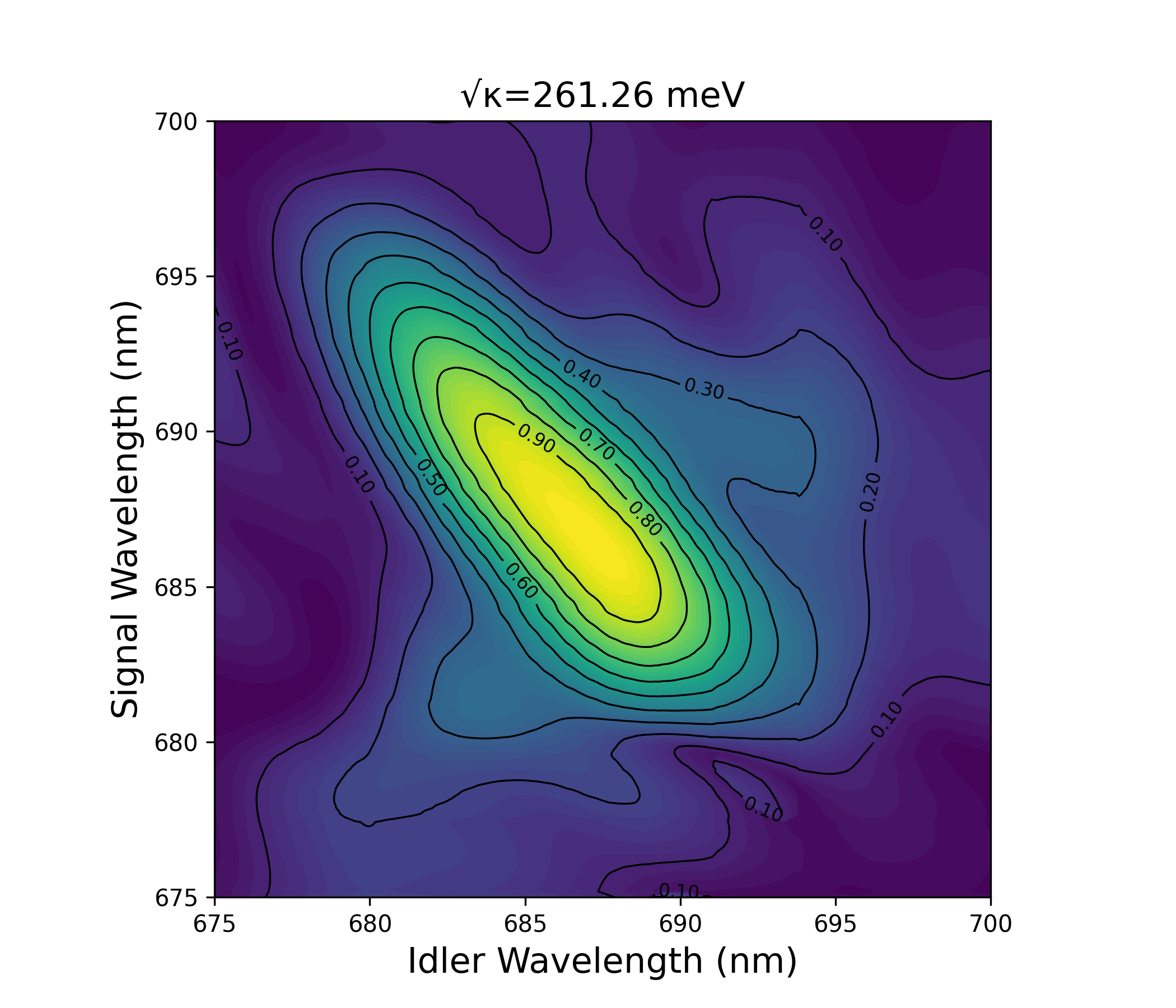}
        \caption{}
        \label{JSI_output_2}
    \end{subfigure}
    \hspace{-0.1\linewidth} 
    \begin{subfigure}[b]{0.45\linewidth}
        \centering
        \includegraphics[width=\textwidth]{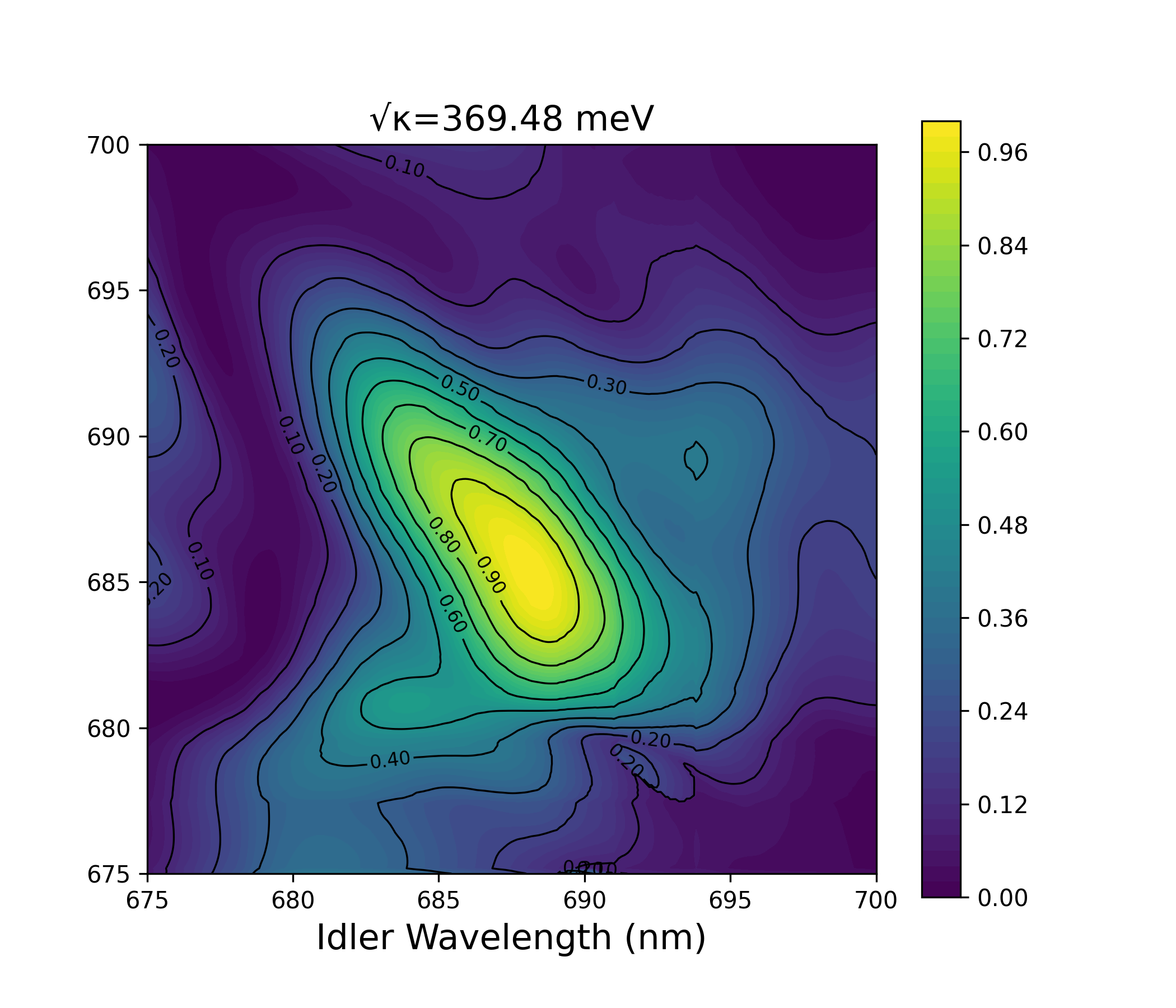}
        \caption{}
        \label{JSI_output_3}
    \end{subfigure}

    \caption{Joint Spectral Intensity (JSI) for different cavity-material coupling strengths $\sqrt{\kappa}$. In (a), we take the experimental input JSI; (b)-(d) show the final JSI of the output state at several cavity-material coupling strengths. We observe the squeezing of the JSI resembling the filtering effect known to occur due to cavities. In addition to expected shift in the peak of the JSI towards higher idler wavelengths, we also see the emergence of off-diagonal peaks. This can be ascribed to the irreversible decay of cavity-material excitations to the output quasi- signal/idler continua in a Lorentzian shape centered at the resonant frequency.}
    \label{JSI_output}
\end{figure*}

 To benchmark the model, we take an experimental input JSI and estimate the dynamics. In Fig. \ref{JSI_output}, we see that the JSI shifts towards higher idler and lower signal wavelengths. Since the system is set up such that the cavity and material dispersions are nearly in resonance with the idler wavelength ($\approx 681nm$), these energies are preferentially absorbed by the cavity and material with a higher probability as compared to the signal photons. As a result, we see a shift in the JSI towards higher idler wavelengths (lower frequencies). For higher coupling strengths, this process occurs at a much faster time scale because of an increase in the Rabi frequency. As a result, for asymptotic output times, the JSI tends to saturate toward higher(lower) idler(signal) wavelengths for higher values of the coupling $\sqrt{\kappa}$. The reciprocal behavior between the signal and idler shifts is due to the conservation of energy. We also see a squeezing of the map, indicating the filtering effect of cavities. This result is in close agreement with the experimental study conducted in empty microcavities \cite{malatesta2023optical} as seen in Fig. \ref{expt_JSI_results}  


\begin{figure*}[!htbp]
    \centering
    \includegraphics[width=\linewidth]{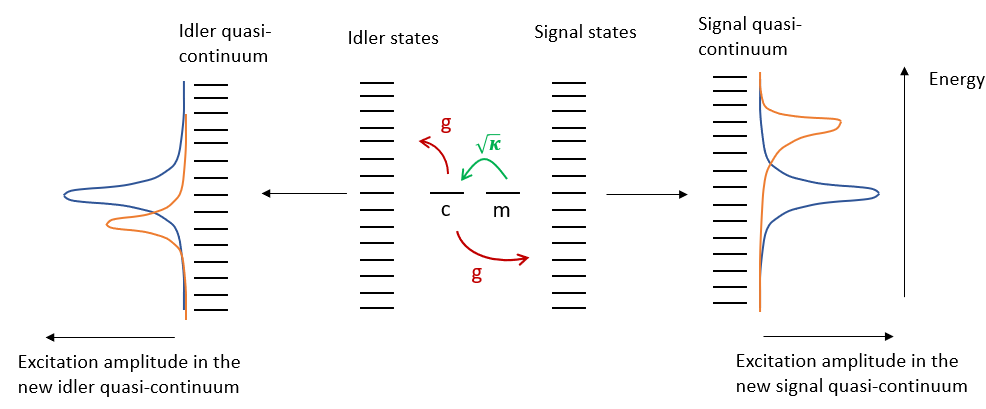}
    \caption{Emergence of peaks along the off-diagonal. The levels ``c'', ``m'' correspond to the cavity and material resonances. The idler and signal continua are discretized in the formalism and presented here as energy levels with constant spacing. The quasi-continua are the states resulting from the coupling of the cavity-material modes with the signal/idler modes. The yellow curves correspond to the initial correlations present in the JSI while the blue curves denote the overlap of the discrete cavity/material states with the quasi-continua. These blue curves are symmetric with respect to the signal and idler channels and thus contribute to the off-diagonal correlations emerging in the output JSI.}
    \label{excitation_decay}
\end{figure*}

\begin{figure}
    \centering
    \includegraphics[width=\linewidth]{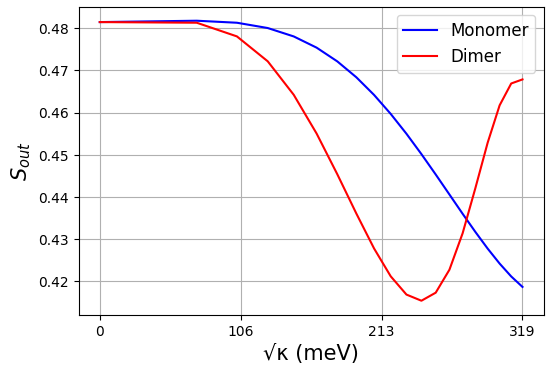}
    \caption{Von Neumann entropy of output for a monomer vs dimer placed within a cavity. We observe a faster decay of entropy for a dimer as compared to that of a monomer.}
    \label{Combined_entropy_output}
\end{figure}

\citet{bittner2020probing} also proved that by measuring the change in the photon entanglement entropy, one can find a direct measure of material correlations. To quantify this, we study its von Neumann entropy given in Eqs. \eqref{svd} and \eqref{vn-entropy}. In Fig. \ref{Combined_entropy_output}, we see that the output entanglement entropy decreases monotonically as a function of the coupling strength, $\sqrt{\kappa}$. This occurs because the relatively preferential absorption of the idler photons by the cavity and material modes degrades the correlations between the signal and idler photons. For dimers, the rate of decrease is higher than that for monomers, as seen by the red curve. This is because excitations can explore a higher Hilbert space and delocalize into these modes faster in the same amount of time. An intriguing aspect of this plot is that the entropy increases after a certain value of the coupling strength. This points toward a revival of strong correlations in the output signal and idler photons.  


Fig.~\ref{JSI_output} also reveals the emergence of peaks along the off-diagonal, indicating an amplification of signal-idler correlations between specific frequency modes. This behavior can be understood by recognizing that our system comprises two discrete modes—the cavity and material excitations—and two continua of states: the signal and idler modes. To capture the dynamics of a discrete mode coupled to a continuum, one typically discretizes the continuum, computes the transition amplitudes, and then takes the continuum limit by letting the discretization spacing approach zero. This procedure enables the calculation of physically relevant quantities, such as level shifts, decay rates, cross sections, and more. A detailed account of this formalism is provided in ~\citet{cohen1998atom}. When a discrete mode couples to multiple discrete states that are well isolated from other states in the system, the resulting transition amplitude becomes a superposition of Rabi oscillations with varying frequencies and amplitudes. In the continuum limit, these superpositions lead to an effective irreversible decay of the discrete mode into the continuum, forming a quasi-continuum. The resulting overlap of the discrete state with the quasi-continuum forms a Lorentzian lineshape centered at the discrete state energy with a width $\hbar\Gamma$, where $\Gamma$ denotes the decay rate of the mode.  

In our setup, when the discrete cavity mode couples to the signal-idler continua, the cavity excitations decay equally into both continua for the same values of signal-idler frequencies. Since the initial correlations are antisymmetric with respect to the signal and idler, and lie along the diagonal, this decay process leads to the emergence of off-diagonal correlations, manifesting as peaks in Fig.~\ref{JSI_output}. The JSI plotted here is obtained by tracing out the signal-idler submatrix from the final correlation matrix of the complete system, and thus reflects the excitations that have decayed into the quasi-continuum.  

Additionally, if the material were coupled to the continua in such a way that the ratio of its couplings did not lie in limiting regimes, the resulting output states would exhibit features characteristic of Fano-type lineshapes. In such a scenario, the JSA lineshapes developed in the output would be asymmetric about the cavity resonance. When applied to both the signal and idler channels, and depending on the detuning between the cavity and material, we expect a superposition of each of these asymmetric lineshapes to appear in the off-diagonal region of the output JSI. These observations would then be characteristic of many-body interactions occuring in our material-cavity system. The trends observed above highlight the capability of photon entanglement to serve as a sensitive probe for exploring many-body interactions and correlations in quantum spectroscopy. 

The simplified structure of our model offers significant analytical tractability and a framework to connect input and output observables. Despite its simplicity, it successfully reproduces the experimentally obtained JSI, and establishes a practical baseline for interpreting the role of entanglement in quantum spectroscopic signals. That said, the formalism inherently excludes certain physical processes, such as pure dephasing, which are typically modeled by $\hat{\sigma}^z$-type jump operators within the Lindbladian. By assuming the Hamiltonian to be in a bilinear form, we preclude the description of non-Markovian dynamics, inter-mode crosstalk, and multi-photon interactions. These effects may arise from energetic disorder, spatial inhomogeneity, or cascaded transitions \cite{li2019photon}. We plan to use the present formalism as a robust foundation and incorporate these complexities in our future investigation.

\section{Summary}
Recent advancements in quantum light spectroscopy have revealed exciting possibilities of using entangled photons as sensitive probes of many-body dynamics and correlations in materials. Although experimental and theoretical efforts continue to evolve in parallel, the literature still lacks a framework that accurately explains observed phenomena. The exponential growth of the Hilbert space and the presence of nonlinear and many-body effects make direct simulations computationally prohibitive. To address this, we reformulated the problem using a finite-sized correlation matrix based on a bilinear bosonic Hamiltonian, leading to Gaussian-preserving time evolutions. This approach allows us to retain the essential physical dynamics that occur in bipolaritonic systems.

We applied our model to an experimentally measured JSI of frequency-entangled biphotons generated through spontaneous parametric down conversion. Our simulations predict a shift in the output JSI that is consistent with the experimental observations. Moreover, we observed the emergence of off-diagonal peaks in the JSI, which indicate enhanced correlations between specific signal-idler frequency pairs. We interpret these features as signatures of a discrete cavity mode decaying into the continua of signal and idler modes. This decay redistributes the spectral amplitudes and gives rise to interference effects similar to Fano resonances. 

These results underscore the sensitivity of biphoton entanglement to serve as a probe in quantum spectroscopic experiments. Finally, we plan to use this as a theoretical baseline to extend and incorporate more complex interactions in our future work.

\begin{acknowledgments}
    The work at the University of Houston was funded by the National Science Foundation (CHE-2404788) and the Robert A. Welch Foundation (E-1337). ARSK acknowledges funding from the National Science Foundation CAREER grant (CHE-2338663), start-up funds from Wake Forest University, funding from the Center for Functional Materials at Wake Forest University. 
\end{acknowledgments}

\section*{Author Contributions}
S.D. developed the theoretical framework, performed the simulations, and wrote the manuscript. A.R.S.K. and E.R.B. supervised the project and contributed to discussions and manuscript revisions. All authors reviewed and approved the final version of the manuscript.

\section*{Data Availability Statement}
All data and source code used in this work are publicly available in the following GitHub repository:
\url{https://github.com/SameerD-phys/Biphoton_entanglement}.

\bibliography{bib-local}

\end{document}